# Multimodal Language Specification for Human Adaptive Mechatronics

Fernando Ferri, Arianna D'Ulizia, Patrizia Grifoni*
*Institute of Research on Population and Social Policies, National Research Council*
*patrizia.grifoni@irpps.cnr.it*
*\*corresponding author*

*Abstract*

*Designing and building automated systems with which people can interact naturally is one of the emerging objective of Mechatronics. In this perspective multimodality and adaptivity represent focal issues, enabling users to communicate more freely and naturally with automated systems. One of the basic problem of multimodal interaction is the fusion process. Current approaches to fusion are mainly two: the former implements the multimodal fusion at dialogue management level, whereas the latter at grammar level. In this paper, we propose a multimodal attribute grammar, that provides constructions both for representing input symbols from different modalities and for modeling semantic and temporal features of multimodal input symbols, enabling the specification of multimodal languages. Moreover, an application of the proposed approach in the context of a multimodal language specification to control a driver assistance system, as robots using different integrated interaction modalities, is given.*

## 1. Introduction

People communicate and interact each other's and with external environment using all the five senses. Hence, a very natural interaction has to involve signals exchanged using all senses in both, voluntary and involuntary communication.

In fact, information that are voluntary conveyed by explicit messages are joined to information provided by involuntary communication, i.e. communication made with one's body (i.e. gestures, eye movement, hand and/or arm placement, facial expression, etc. not explicitly conceived by the user for communicating with others or the environment).

The importance of voluntary and involuntary communication in the interaction processes between humans, humans and environment, humans and mechatronic systems can be easily taken off. Indeed, naturalness in the interaction process requires that the human and the mechatronic system have the same autonomy level removing the user's control function of the first on the latter.

Mechatronics integrates mechanical, electronical and information-driven units allowed for turning conventionally designed mechanical components into smart devices.

In particular, mechatronic systems can exchange messages among them and with people and/or they can percept environmental changes using multimodal input and output actions. Recently, several authors have analyzed principles, features, and challenges of multimodal interaction [1] [2] and how it benefits the disambiguation of the dialogue [3] as well as the cognitive and learning processes of human-machine systems [4]. In [5], these principles, features, and challenges have been discussed in the particular perspective of mobile devices.

In most of the existing human-machine systems, humans have to learn, to act on machines and to acquire skills by themselves, without the machines' assistance.

Considering actions as part of the communication process (i.e. from the events produced by someone on some other in order to modify it, to a reciprocal conversation between two or more entities) to improve the interaction effectiveness and naturalness between a human and a machine, the mechatronics should focus on human adapting the system's behaviour to the operator's skill level. That is, the Human Adaptive Mechatronics (HAM) represents a new and relevant challenge to design mechatronic systems. In this perspective, Harashima et al. [6] identified some relevant issues, such as:

*"Modeling human and machine dynamics. Especially the variable constraints should be considered. Modeling the operation base on skill, rule, knowledge, decision combining event and dynamical*

*systems. Modeling of the psycho-physical characteristics of human operator. Development of mechatronic systems supporting human operation not only assisting control action but also providing proper data and knowledge for understanding the situation and giving better decision."*

Such a mechatronics has to be equipped with abilities to observe human behaviour and environment events and create actions. Modelling human-machine dynamics includes the development of an interaction language that should be a multimodal language (which will be introduced in Section 3), since multimodal interaction combines and integrates information from different input modalities (fusion process), and generates appropriate output information (fission process) enabling a natural dialogue. An integrated and general software environment for multimodal interaction languages specification is proposed to enhance and adapt the mechatronics' interaction to the user.

We assume that programming, cooperating and interacting with a mechatronic system involve multimodal inputs, which means that the user can program the system "by example" by simply using speech, gesture or other modalities opportunely combined. Moreover, the system can acquire and update the multimodal interaction language according to the different contexts and users, where context refers to everything characterising the system interaction process with humans, according to the notion provided by Dey [7].

In particular, in this paper we focus our attention on defining a grammar-based approach for specifying and updating a multimodal language for interacting with a mechatronic system. For this purpose, we propose a multimodal attribute grammar, that provides constructions both for representing input symbols from different modalities and for modeling semantic and temporal aspects of multimodal dialogue. Specifically, we start from attribute grammars, firstly defined by Donald Knuth [8] as a means of formalizing the semantics of a context-free language[1], and we extended these grammars introducing a multimodal attribute grammar that provides constructions both for representing input symbols from different modalities and for modeling semantic and temporal aspects of the multimodal dialogue.

The remainder of the paper is structured as follows. Section 2 briefly describes research activities related to this work focusing on mechatronics and human-machine interaction. In Section 3 a preliminary analysis of challenges and existing approaches to multimodal dialog interpretation is provided and an overview of our theoretical approach based on a multimodal attribute grammar is given. Section 4 illustrates the current implementation of the proposed approach applied to a driver assistance system that allows speech-gesture interaction. Section 5 gives experimental details and results of the tests that we performed to validate the proposed approach. Finally, conclusions and future work are given in Section 6.

## 2. Related work

In the last years, the use of mechatronics is moving from the industrial sector towards the daily living environments, causing the need of assuring a fruitful interaction between humans and these systems. With the aim of responding to this need, the research on Human Adaptive Mechatronics (HAM) has emerged as an interdisciplinary research field aiming at designing and developing systems capable of adapting their behavior to the human abilities. HAM is a concept firstly introduced at the Tokyo Denki University in the COE (Center of Excellence) research project (sponsored by Japanese Ministry of Education, Sports, Culture, Science and Technology), and it is defined as "intelligent mechanical systems that adapt themselves to the user's skill under various environments, assist to improve the user's skill, and assist the human-machine system to achieve best performance".

According to these goals, a mechatronic system has to be able to interact with humans like it will be a human too, enhancing the system's role in a peer-to-peer human-system interaction approach as in [9].

Usually, mechatronic systems use predefined interaction language specifically oriented to a particular application; an interesting experience among them is described by Iba et al. [10]. They use multimodal interaction between humans and robots in order to describe a framework to compose robot programs by non expert people. Two key ideas of this programming approach are that the user can interactively provide a feedback at any time through an intuitive interface and that the system infers the user's intent to support interaction. The programs produced by the system have the limit that are composed by fixed task sequences.

---

[1] A context-free language is a language over some alphabet that can be generated from a particular kind of a grammar known as a context-free grammar.

A more general approach has been adopted in [11], where the authors propose a software environment for adaptively acquiring a multimodal language during multimodal interaction with a complex environment. In particular, they investigate language acquisition from spoken input combined with simultaneous mouse or pen input. The proposed approach is independent from such a particular context, but it is limited in the interaction because the language acquisition was carried out by spoken input combined with simultaneous mouse or pen input only.

Wermter et al. [12] have studied the problem of learning in intelligent robots using an integrated approach based on demonstration and imitation. In particular, motor actions, vision and language instructions are integrated by means of both a self-organizing network, which associates the action sensor readings with the action verbs, and an associator neural network for localizing the recognized object within the visual field.

Our work, differently from the previous ones, is devoted to specify how to define a more general multimodal language and how humans and a mechatronic system can interact using it.

## 3. A grammar-based approach for the multimodal dialog interpretation

Multimodal interaction allows the user to communicate with mechatronic systems through the simultaneous or alternative use of several channels of input/output at a time. Such a kind of interaction offers a more flexible, efficient and usable communication way allowing the user to interact through input modalities, such as speech, handwriting, hand gesture and gaze, and to receive information by the system through output modalities, such as speech synthesis and smart graphics.

In multimodal interaction the two main challenges to face are: to combine, integrate, and interpret information from different input modalities (fusion process), and to generate appropriate output information (fission process) in order to enable a natural dialogue between users and mechatronic systems. Our specific concern in this paper is with the integration and interpretation of multiple input modalities.

### 3.1. Challenges and approaches to the multimodal dialog interpretation

In the literature, two different approaches to the fusion process have been proposed. The first one, which we refer to as grammar-based approach, combines the multimodal inputs at grammar level. This means that the different unimodal inputs are considered as a unique multimodal input by using the multimodal grammar specification. Subsequently, the dialogue parser applies the grammar rules to interpret the multimodal sentence. The second strategy, which we refer to as dialogue-based approach, combines the multimodal inputs at dialogue level. This means that the different unimodal inputs are distinctly interpreted and then they are combined by the dialogue management system.

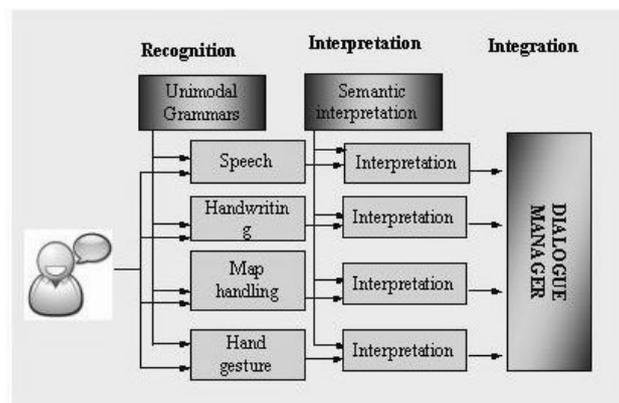

**Figure 1.** A dialogue-based fusion system

In multimodal systems, both dialogue and grammar based, it is common to have separate recognizers for each modality managed by the system. This means that the recognition phase (see Figure 1 and 2) is the same for the two approaches. The difference is in the interpretation and integration phases.

In dialogue-based fusion, the outcomes of each recognizer are separately interpreted and then sent to the dialogue manager that performs their fusion by using integration mechanisms, such as, for example, statistical integration techniques, agent theory, hidden Markov models, artificial neural networks. Figure 1 illustrates a general architecture for this strategy.

This approach was followed by several authors in the literature. Corradini et al. [13] proposed a dialogue-based approach for modality fusion in which the fusion engine combines the time-stamped information received from the recognizers, selects the most probable multimodal interpretation, and passes it to the dialogue manager. The selection of the most probable interpretation is carried out by the dialogue manager that rules out inconsistent information by both binding the semantic attributes of different modalities and using environment content to disambiguate information from the single modalities. Other two dialogue-based approaches are ICARE [14] and MIMUS [15]. The former considers both pure modalities, described through elementary components, and combined modalities, specified by the designer through composition components. The fusion is performed within the dialogue manager by using a technique based on agents (PAC agents). The MIMUS approach starts from the assumption that each individual input can be considered as an independent dialogue move. If more than one input is received during a certain time frame, the dialogue manager performs further analysis in order to determine whether those independent multimodal inputs are truly related or not.

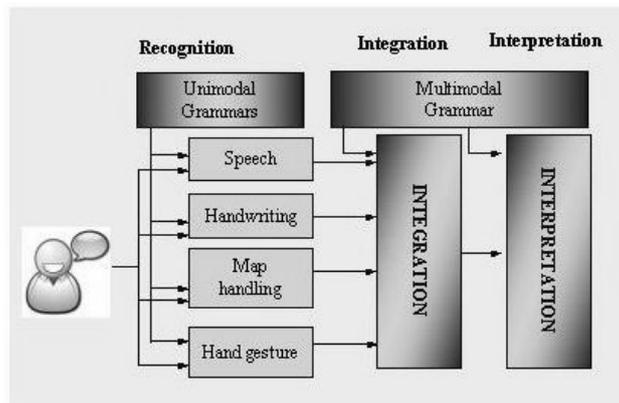

**Figure 2.** A grammar-based fusion system

In the grammar-based approach, the outcomes of each recognizer are considered as terminal symbols of a formal grammar and consequently they are recognized by the parser as a unique multimodal sentence. Therefore, the fusion of the multimodal inputs occurs at grammar level. Finally, in the interpretation phase the parser uses the grammar specification (production rules) to interpret the sentence. Figure 2 illustrates a general architecture for this strategy.

The grammar-based approach was followed by Johnston and Bangalore [16]. The authors perform multimodal parsing and understanding by using a weighted finite-state machine. Modality integration is carried out by merging and encoding into a finite-state device both semantic and syntactic content from multiple streams. In this way, the structure and the interpretation of multimodal utterances can be captured declaratively in a context-free multimodal grammar. Another grammar-based approach is MUMIF [17]. The fusion module of MUMIF applies a multimodal grammar to unify the recognized unimodal inputs into a unique multimodal input that is represented by using the TFS (Typed Feature Structures) structure proposed by Carpenter [18]. The MUMIF multimodal grammar is composed of two kinds of rules: lexical rules that are used to specify the TFS representation and grammar rules that constrain the unification of input. D'Ulizia et al. [19] [20] [21] also followed the grammar-based approach and proposed a hybrid grammar that provides a logical level over a multi-dimensional grammar. Specifically, the authors uses a constraints multi-set grammar (CMG) for representing the structure of multimodal inputs and logical rules for specifying the temporal and semantic constraints of multimodal inputs.

A comparison of dialogue-based and grammar-based fusion strategies has been made by Manchón et al. [22], which concluded that the grammar-based approach is the most natural one as it is more coherent with the human-human communication paradigm, in which the dialogue is seen as a unique and multimodal communication act. Moreover, this approach allows an easier inter-modality disambiguation.

Therefore, the use of a grammar, for supporting the mechatronics in parsing and interpreting the multimodal sentences expressed by the user, enables a more flexible and natural interaction. Since a large number of grammars has been defined for natural language processing, they represent a valuable and standardized starting point toward the extension to multimodal input. In this paper, we propose a multimodal attribute grammar, which extends the attribute grammars firstly defined by Donald Knuth [8]. The proposed grammatical formalism provides constructions for representing input symbols from different modalities (and how they have to be combined together into the input sentence), and for modeling semantic and temporal aspects of input symbols, as described in the following section.

**3.2. The proposed grammatical formalism**

A promising approach for defining grammars for multimodal languages consists in starting from techniques used in Natural Language Processing (NLP) (see [23]) and extending them to Multimodal Language Processing (MLP). As traditional grammars for natural language (NL) (that is, the kind of language used by human beings) are not powerful enough to cope with the syntactic structure of multimodal languages, an evolution of NL grammars towards multimodal grammar is necessary.

The multimodal grammar proposed in this paper follows the context-free paradigm due to its ability to model all frequent linguistic constructions of multimodal language by assuring, at the same time, a lower parsing complexity. However, in order to use context-free grammars (CFG)[2] for MLP it is necessary to overcome the two main deficiencies of this grammatical formalism, i.e. the lack of constructions both for representing input symbols from different modalities and for modeling semantic and temporal aspects of input symbols.

In this attempt, attribute grammars provide a good compromise between the context-free paradigm and the necessity to represent semantic and temporal aspects of multimodal input.

Attribute grammars [8] were firstly developed as a means of formalizing the semantics of a context-free language. They may be informally defined as a context-free grammar that has been extended to provide context sensitivity using a set of attributes (associated with each distinct symbol in the grammar), assignment of attribute values, evaluation rules, and conditions.

Starting from the attribute grammar formalism, an extension of this notation for multimodal input processing is necessary. Therefore, the definition of Multimodal Attribute Grammar (MAG), described in [24], has been introduced and given synthetically below.

**Definition 1.** *A Multimodal Attribute Grammar is a triple*

$$G = (\mathcal{G}, \mathcal{A}, \mathcal{R})$$

*where:*

*(1) $\mathcal{G}$ is a context-free grammar (T,N,P,S) with T as set of terminal symbols, N as set of non-terminal symbols, P as set of production rules of the form:*

$X_0 \to X_1 X_2 \ldots X_n$     *where $n \geq 1$, $X_0 \in N$ and $X_k \in N \cup T$ for $1 \leq k \leq n$*

*and $S \in N$ as start symbol (or axiom)*

---

[2] A context-free grammar (CFG) is a formal grammar $G = (N, \Sigma, P, S)$ in which every production rule $p \in P$ is of the form $A \to w$, where $A \in N$ (i.e. $A$ is a non-terminal symbol) and $w \in (N \cup \Sigma)^*$ (i.e. $w$ is a string consisting of terminal and/or non-terminal symbols).

*(2) $\mathcal{A}$ is a collection $(A(X))_{X \in N \cup T}$ of attributes of the non-terminal and terminal symbols, such that for each $X \in N \cup T$, $A(X)$ is split in two finite disjoint subsets $I(X)$, the set of inherited attributes of $X$, and $S(X)$, the set of synthesized attributes. The set $S(X)$ with $X \in T$ includes a set of attributes $MS(X)$, called set of multimodal synthesized attributes, composed of the following four attributes:*

$$MS(X) = \{val, mod, synrole, coop\}$$

*(3) $\mathcal{R}$ is a collection $(R_p)_{p \in P}$ of semantic functions (or rules)* ∎

The attributes of the set $MS(X)$ are very general and independent from the application domain. They manage the multimodal properties of the sentence's symbols: value, modality, syntactic role, and modality cooperation type. This information is represented into the four attributes of $MS(X)$:

- *val* that expresses the current value (concept) of the terminal symbol. The domain of the attribute is the set of terminal symbols: $D_{val} = T$
- *mod* that represents the modality. The domain of the attribute is the set of modalities (in our system we have four modalities): $D_{mod} = \{speech, handwriting, gesture, sketch\}$
- *synrole* that conveys information about the syntactic role. The domain of the attribute is $D_{synrole} = \{noun\ phrase, verb\ phrase, determiner, verb, noun, adjective, preposition, deictic, conjunction\}$
- *coop* that expresses the modality cooperation type with other terminal symbols. The domain of the attribute is $D_{coop} = \{complementary, redundant\}$

The set $R_p$ of semantic functions, associated with each production rule $p$ in $P$, allows to compute the values of the attributes. They are of the form:

$$X_i.b \leftarrow f(y_1, ..., y_k) \qquad \text{with } k \geq 1$$

where

1. $X_i.b$ denotes the occurrence of the attribute $b$ of the symbol $X_i$ in the production rule $p \in P$;

2. $y_j$, with $1 \leq j \leq k$, is an occurrence of the attribute of a symbol in the body of the production rule $p \in P$;

3. $f$ is a function that maps the values of $y_1, ..., y_k$ to the value of $X_i.b$.

The strength of the MAG is the capability to manage whatever modalities and to represent temporal constraints into the grammar rules. Moreover, it provides a good compromise between the context-free paradigm and the necessity to represent semantic and temporal aspects of multimodal input.

The process of adding new information to the basic ground of the mechatronic system is essential in order to recognize not only very well defined sentences but also different sentences with a similar semantic meaning. This process requires the runtime updating of the grammar. To achieve that we use the algorithm proposed in [25] that is inspired by the Cocke-Younger-Kasami (CYK) algorithm [26]. In particular, if the multimodal sentence expressed by the user is not inferable from the set of production rules of the current grammar, our algorithm generates the minimum set of production rules and adds them to the existing set of production rules under the control of the user that verifies the semantic meaning of the sentence. Therefore, the multimodal grammar is continuously updated by generating production rules in order to include and recognize various multimodal sentences.

## 4. The driver assistance system

The main focus at this stage of our study is to apply the proposed grammatical formalism to a mechatronic system in order to illustrate the current implementation and preliminary results of the method. In particular, the applicative domain that we have simulated is a driver assistance system that is equipped with mobile components, communication facilities, and user interaction possibilities.

Generally, driver assistance systems have three main functionalities to support the driver:
- informative, as they can provide traffic, meteorology and navigational information;
- entertainment, as they can provide multimedia applications for listening to music or seeing videos;

- safety control, as they can actively support the driver performing driving tasks by reducing and/or compensating errors of the driver.

A driver assistance system is considered a mechatronic system as it is made of mechanical, electrical, and information technology systems and control engineering. In fact, mainly the safety control functionalities use sensors, cameras and computer systems to detect the road condition and to consequently intervene with the driving task by using throttle, brake or steering wheel controls.

A higher cooperation between the assistance system and the driver is needed, so that this system is a really safe and effective driving tool. In particular, a multimodal interaction is essential for communicating with the system without distracting from the driving task.

### 4.1. The Architecture

In our specific application, the user can ask the system: questions related to the car state and the traffic condition, to set some driving options, to switch lights on or off, to call emergency or breakdown services. The user communicates with the system by voice and gesture. The gesture modality refers to the pointing of an object (icon, map, text, etc) on a touch screen by the hand.

The current software components of this driver assistance system include a speech and gesture recognizers, a multimodal fusion component, and a speech synthesizer.

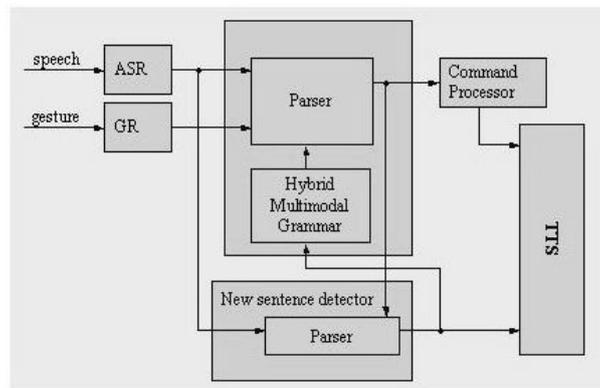

**Figure 3.** The architecture of the multimodal dialogue component

The architecture of the multimodal dialogue component of the system is given in Figure 3. We predisposed speech and gesture recognition modules that give as output the utterances of each recognized spoken word and the meaning of each recognized gesture, respectively. In particular, the function of the ASR is to translate incoming speech signal into a structured stream of words, each one having an associated set of attributes composed of the concept represented by the word and the start/end times. Analogously, the GR deals with the translation of hand gestures into a structured stream of gesture symbols, each one having an associated set of attributes like that of spoken words. These outputs are the terminal symbols of our multimodal attribute grammar (see Section 4.2).

Initially, the multimodal grammar contains a set of production rules that allows the interpretation of a set of multimodal sentences, each one corresponding to a command for the mechatronic system. If the user's multimodal sentence is not interpretable by using the available production rules, the parser sends the sentence to the "New sentence detector" module. This module asks the user by synthetic voice to provide a meaning of the sentence, that is the expected action of the system. The algorithm for the incremental learning of the grammar is applied and generates the set of production rules to parse the multimodal sentence. Consequently, the production rules of the hybrid multimodal grammar are updated in order to allow also the interpretation of the new sentence. The whole application was written in Java.

### 4.2. The grammar for the driver assistance system

The driver assistance system allows the user to interact multimodally in order to ask information about the car state or the traffic condition, to set some driving options, to switch lights on/off, to call emergency or breakdown services. To achieve that, a multimodal attribute grammar is predisposed containing a set of 26 production rules, shown in Table 1, and the following sets of terminal and non-terminal symbols:

T = {*Call, Save, Recall, Delete, Play, Help, Turn off, Repeat, Read, Turn on, Person, Number, Phone-book, Song, CD, Station, Temperature, Defrost, USB, This*}
N = {*NOUN, VP, VERBT, NP, DT, S*}

**Table 1.** The production rules and semantic functions of the multimodal attribute grammar for the driver assistance system

```
P1) S → NOUN                         P11) DT → This                            P19) NOUN → station
    R1.1) S.val ← NOUN.val                R11.1) DT.val ← this                      R19.1) NOUN.val ← station
    R1.2) S.mod ← NOUN.mod                R11.2) DT.mod ← speech                    R19.2) NOUN.mod ← speech | gesture
                                          R11.3) DT.synrole ← deictic               R19.3) NOUN.synrole ← noun
P2) S → VP                                R11.4) DT.coop ← complementary
    R2.1) S.val ← VP.val                                                        P20) VERBT → Turn off
    R2.2) S.mod ← VP.mod             P12) VERBT → Play                              R20.1) VERBT.val ← turn off
                                          R12.1) VERBT.val ← play                   R20.2) VERBT.mod ← speech
P3) VP → VERBT                            R12.2) VERBT.mod ← speech                 R20.3) VERBT.synrole ← verb
    R3.1) VP.val ← VERBT.val              R12.3) VERBT.synrole ← verb
    R3.2) VP.mod ← VERBT.mod                                                    P21) NOUN → Temperature
                                     P13) NOUN → Help                               R21.1) NOUN.val ← temperature
P4) VP → VERBT NP                         R13.1) NOUN.val ← help                    R21.2) NOUN.mod ← speech | gesture
    R4.1) VP.val ← VERBT.val + NP.val     R13.2) NOUN.mod ← speech | gesture        R21.3) NOUN.synrole ← noun
    R4.2) VP.mod ← VERBT.mod + NP.mod     R13.3) NOUN.synrole ← noun
                                                                                P22) NOUN → Defrost
P5) NP → DT NOUN                     P14) NOUN → Person                             R22.1) NOUN.val ← defrost
    R5.1) NP.val ← NOUN.val               R14.1) NOUN.val ← person                  R22.2) NOUN.mod ← speech | gesture
    R5.2) NP.mod ← DT.mod + NOUN.mod      R14.2) NOUN.mod ← speech                  R22.3) NOUN.synrole ← noun
                                          R14.3) NOUN.synrole ← noun
P6) NP → NOUN                             R14.4) NOUN.coop ← complementary     P23) NOUN → USB
    R6.1) NP.val ← NOUN.val                                                         R23.1) NOUN.val ← usb
    R6.2) NP.mod ← NOUN.mod          P15) NOUN → Number                              R23.2) NOUN.mod ← speech
                                          R15.1) NOUN.val ← number                  R23.3) NOUN.synrole ← noun
P7) VERBT → Save                          R15.2) NOUN.mod ← speech | gesture
    R7.1) VERBT.val ← save                R15.3) NOUN.synrole ← noun           P24) VERBT → Repeat
    R7.2) VERBT.mod ← speech                                                        R24.1) VERBT.val ← repeat
    R7.3) VERBT.synrole ← verb       P16) NOUN → Phone-book                          R24.2) VERBT.mod ← speech
                                          R16.1) NOUN.val ← phoe-book               R24.3) VERBT.synrole ← verb
P8) VERBT → Call                          R16.2) NOUN.mod ← speech | gesture
    R8.1) VERBT.val ← call                R16.3) NOUN.synrole ← noun           P25) VERBT → Read
    R8.2) VERBT.mod ← speech                                                        R25.1) VERBT.val ← read
    R8.3) VERBT.synrole ← verb       P17) NOUN → Song                                R25.2) VERBT.mod ← speech
                                          R17.1) NOUN.val ← song                    R25.3) VERBT.synrole ← verb
P9) VERBT → Recall                        R17.2) NOUN.mod ← speech | gesture
    R9.1) VERBT.val ← recall              R17.3) NOUN.synrole ← noun           P26) VERBT → Turn on
    R9.2) VERBT.mod ← speech                                                        R26.1) VERBT.val ← turn on
    R9.3) VERBT.synrole ← verb       P18) NOUN → CD                                  R26.2) VERBT.mod ← speech
                                          R18.1) NOUN.val ← cd                      R26.3) VERBT.synrole ← verb
P10) VERBT → Delete                       R18.2) NOUN.mod ← speech | gesture
    R10.1) VERBT.val ← delete             R18.3) NOUN.synrole ← noun
    R10.2) VERBT.mod ← speech
    R10.3) VERBT.synrole ← verb
```

The set of production rules is generated "by example" using the interface shown in Figure 4, where the multimodal sentences that have to be recognized are inserted and the system applies the algorithm for the incremental learning of the grammar that generates the basic set of production rules, visualized in the right side of Figure 4.

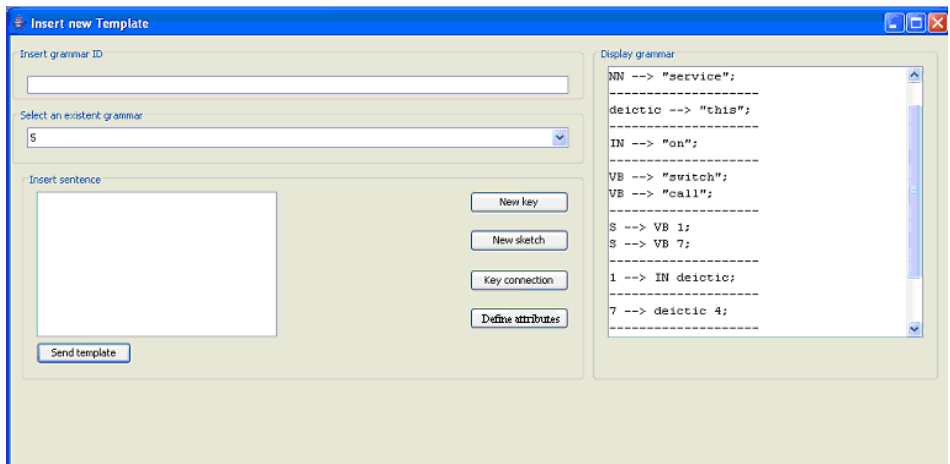

**Figure 4.** The interface for the definition "by example" of the basic set of production rules of the multimodal grammar

Examples of multimodal sentences that have to be recognized are the following:

{S1: speech: "Call this breakdown service"
gesture: to indicate the breakdown service on a touch-screen display
S2: speech: "Turn on/off this"
gesture: to point the temperature icon
S3: speech: "Play this song"
gesture: to point the name of the song on a touch-screen display
S4: speech: "Save this station"
gesture: to indicate the number of the radio station}

## 5. Evaluation

In order to assess the usability of the implemented driver assistance system, some experiments were conducted. The objective of these experiments was to evaluate the system qualitatively and quantitatively by collecting subjective usability measures about the satisfaction, ease-of-use, user-friendliness, and time.

The total number of participants in the experiment was twelve. These people have been divided in two groups: group 1 (three males and three females) involved people with high skill with multimodal languages, and group 2 (three males and three females) was composed of people without any skill with multimodal languages.

For conducting the experiments we asked participants to interact with the driver assistance system we predisposed. The experimental procedure included two interactive tasks. The former requires the user to switch lights on by expressing the multimodal sentence S5 (speech: "Turn on this" + gesture: to point the headlight icon), that is included in the basic set of multimodal sentences recognizable by the system. The latter requires the user to set the speaker volume by expressing the multimodal sentence S6 (speech: "set to 15" + gesture: to point the speaker volume icon), which is not included in the basic set of multimodal sentences recognizable by the system.

For each task, participants are preliminarily instructed to use the driver assistance system by watching an instructional video.

The criteria we used for evaluating the performance of the system are the following:
- the task is successfully completed or not;
- the grammar is correctly generated;
- which is the satisfaction of the user.

For evaluating the system according to the first criteria we asked participants to accomplish the two tasks and we count the total number of tasks all participants carried out.

For evaluating if the grammar is correctly generated, participants expressed the multimodal sentence S6 and we evaluate if the production rules generated by the incremental learning algorithm are able to parse it. We count the total number of parsed sentences.

The third evaluation criteria consists in a final questionnaire in which we asked participants to grade the performance of the system, on a 5-point Likert scale ranging from "strongly disagree" to "strongly agree". A summary of the evaluation results we obtained is illustrated in Table 2. The results of the experiments demonstrate that the approach ensures a good multimodal accuracy.

**Table 2.** Experiment results of the approach

| | |
|---|---|
| Total number of accomplished tasks | 79,1% |
| Total number of parsed sentences | 83,3% |
| Satisfaction of the user | 69,8% |

## 6. Conclusion and future work

In this paper, we presented a novel approach to multimodal dialog definition and interpretation, which is based on a multimodal attribute grammar that has the strength to manage whatever modalities and to represent temporal constraints into the grammar rules. In particular, we started from attribute grammars, firstly defined by Donald Knuth [8] as a means of formalizing the semantics of a context-free language, and we extended these grammars introducing constructions both for representing input symbols from different modalities and for modeling semantic and temporal aspects of input symbols.

The advantages of this new grammatical formalism are that it allows the reasoning about multimodal languages providing a prerequisite for being able to resolve ambiguities and for integrating parsing with semantic theories. Moreover, the algorithm for the incremental learning of the grammar allows adding new information to the basic ground of the mechatronic system enabling the recognition of not only very well defined sentences but also different sentences with a similar semantic meaning.

Therefore, we applied this theoretical approach for the definition and interpretation of the dialog between a human and a particular kind of mechatronics, that is a driver assistance system. The results we obtained from the evaluation phase demonstrate that the approach ensures a good multimodal accuracy.

Future work would involve the development of a complete framework based on the multimodal attribute grammar, which allows the formalization of multimodal interaction languages. Moreover, we would improve the evaluation phase by enlarging the number of people involved in the experiments in order to calculate the effectiveness of our approach and to compare it to other existing approaches.